# Room-temperature ferromagnetism in nanoparticles of Superconducting materials

**Shipra, A. Gomathi, A. Sundaresan\* and C. N. R. Rao\***

*Chemistry and Physics of Materials Unit and*
*Department of Science and Technology Unit on Nanoscience,*
*Jawaharlal Nehru Centre for Advanced Scientific Research,*
*Jakkur P. O., Bangalore 560 064 India*

**Abstract**

Nanoparticles of superconducting $YBa_2Cu_3O_{7-\delta}$ (YBCO) ($T_c$ = 91 K) exhibit ferromagnetism at room temperature while the bulk YBCO, obtained by heating the nanoparticles at high temperature (940$^o$ C), shows a linear magnetization curve. Across the superconducting transition temperature, the magnetization curve changes from that of a soft ferromagnet to a superconductor. Furthermore, our experiments reveal that not only nanoparticles of metal oxides but also metal nitrides such as NbN ($T_c$ = 6 - 12 K) and δ-MoN ($T_c \sim$ 6 K) exhibit room-temperature ferromagnetism.

\*Electronic address: sundaresan@jncasr.ac.in
\*Electronic address: cnrrao@jncasr.ac.in

**1. Introduction**

It is well known that the materials with partially filled *d* or *f* shells exhibit interesting magnetic properties. In the recent years, ferromagnetism above room temperature has been observed in several materials containing no unpaired *d* or *f* electrons. High-temperature ferromagnetism has been reported in alkaline-earth hexaboride.[1] Ferromagnetic hysteresis and high Curie temperature was observed in proton irradiated graphite.[2] Nanoparticles of gold capped with thiol were shown to exhibit ferromagnetism at room temperature.[3] Thin films of many oxides with no unpaired electrons are also reported to be ferromagnetic at room temperature.[4,5] It was suggested that defects or oxygen vacancies could be responsible for the observed ferromagnetism. Ab *initio* electronic structure calculations using density functional theory in $HfO_2$ have shown that isolated halfnium vacancies lead to ferromagnetism.[6] Meanwhile, there is a conflicting report attributing the ferromagnetism in $HfO_2$ to possible iron contamination while using stainless-steel tweezers in handling thin films.[7]

Very recently, we have reported room-temperature ferromagnetism in nanoparticles of nonmagnetic oxides such as $CeO_2$ and $Al_2O_3$.[8] The origin of ferromagnetism in these nanoparticles has been suggested to be due to magnetic moments arising from oxygen vacancies at the surfaces of the nanoparticles. We also suggested that all oxide nanoparticles would exhibit ferromagnetism at room temperature. These results motivated us to study whether the ferromagnetism occurs in nanoparticles of superconductors in the normal as well as the superconducting states, since the phenomena of ferromagnetism and superconductivity are antagonistic. For this purpose, we selected the well known oxide high temperature superconductor, $YBa_2Cu_3O_{7-\delta}$ (YBCO). We also examined whether the ferromagnetism exists in the nanoparticles of metal nitrides such as NbN and δ-MoN which are also known to be superconductors with low $T_c$. Interestingly, we found room-temperature ferromagnetism not only in the nanoparticles of oxide superconductor but also in the metal nitrides.

## 2. Experimental

Nanoparticles of YBCO were prepared by the citrate-gel method described in the literature except that the sample was cooled from 790$^o$ C to room temperature at a faster rate (5$^o$ C/min) than the reported one (1$^o$ C/min).[9] The nanoparticles of NbN were prepared by heating $NbCl_5$ with urea in 1:6 molar ratio at ~ 1000$^o$ C.[10] The nanoparticles of δ-MoN were prepared by heating γ-$Mo_2N$ nanoparticles which in turn were prepared by the urea method described elsewhere.[11] Powder x-ray diffraction (XRD) was used to identify the phase and its purity. The particle size and morphology were studied by Field Emission Scanning Electron Microscopy (FESEM) and Transmission Electron Microscopy (TEM). Magnetization measurements were carried out with vibrating sample magnetometer in physical property measuring system (PPMS, Quantum Design, USA).

## 3. Results and discussion

The XRD pattern of nanoparticles of YBCO showed the orthorhombic structure with a = 3.829 Å, b = 3.874 Å and c = 11.670 Å. The XRD pattern of NbN nanoparticles was consistent with the cubic lattice with a = 4.44 Å. The structure of δ-MoN was confirmed to be hexagonal with a = 5.68 Å and c = 5.56 Å. The particle size of YBCO ranges from 100 - 200 nm as seen from the image of FESEM (Fig. 1). The particles size of δ-MoN and NbN were 20 and 30 nm, respectively.[10]

Fig. 2 shows zero-field cooled (ZFC) and field cooled (FC) susceptibility data as a function of temperature. It can be seen that nanoparticles of YBCO exhibit superconducting diamagnetism with a critical temperature, $T_c$ = 91 K similar to that of bulk YBCO despite the faster cooling rate adopted in this work. In contrast to bulk YBCO, the reversible region in the nanoparticles extends down to 40 K. This is due the increased threshold for flux penetration in the nanoparticles due to Bean-Livingston surface barrier.[12] The magnetization versus field curve measured at 5 K is shown in the inset of the Fig. 2. The field decreasing part of the magnetization remains negative, consistent with the granular nature of the superconductor where the flux pinning is also weak compared to the bulk or single crystals.

Intriguingly, the room-temperature magnetization of YBCO nanoparticles show hysteresis, with a coercivity of ~ 200 Oe, typical of ferromagnetic behavior as shown in Fig. 3. This observation is consistent with our prediction that all oxide nanoparticles would exhibit ferromagnetism.[8] The origin of ferromagnetism is likely to be due to magnetic moments arising from the oxygen vacancies at the surfaces of the nanoparticles. A bulk YBCO was prepared by pressing the nanoparticles into rectangular bars and annealed at high temperatures. The magnetization measurement at room temperature revealed paramagnetic behavior (Fig. 3) which is typical of bulk YBCO. This supports the suggestion that the ferromagnetism is confined to the surface of the nanoparticles. It would be interesting to explore whether such surface ferromagnetism exists in the superconducting state of the nanoparticles. For this purpose we measured M(H) across the superconducting transition. At $T_c$ (= 91 K), the ferromagnetic hysteresis still remains with increased coercivity (300 Oe). At 90 K, the M(H) data is shown in the inset of Fig. 3. It can be seen that the M(H) behavior is that of a superconductor rather than that of ferromagnetism. In fact, it would be difficult to infer the presence of ferromagnetism below $T_c$ from the magnetization measurement.

NbN nanoparticles showed $T_c$ in the range 6 - 12 K depending on the preparative conditions. The diamagnetic susceptibility and M(H) curve in the superconducting state are shown in the inset of Fig. 4. More importantly, they also exhibit room-temperature ferromagnetism with coercivity ~ 500 Oe as seen in Fig. 4. In order to compare this result with bulk or bigger particles of NbN, the nanoparticles were pressed into bars and heated at high temperature (920$^o$C) for a prolonged time in ammonia. This treatment gave an average particle size of ~ 200 nm. The results of the magnetic measurement are shown in Fig. 5. It can be seen that the magnetic hysteresis measured in the superconducting state (T=3.3 K) is that of a bulk sample. Further, the magnetization measured at room-temperature shows essentially a linear behavior but for a small hysteresis and step at very low fields which arises from the tape used to fix the sample with the quartz sample holder. However, it is remarkable to note the M(H) behavior of smaller (Fig. 4) and bigger (Fig. 5) particles of NbN which demonstrate that the ferromagnetism is associated with the smaller particles. Further, as there was no oxide impurity phase in the x-ray diffraction, we suggest that the observed ferromagnetism is intrinsic to the nanoparticles of NbN. This suggests that there can be vacancies of nitrogen at the surfaces of the nanoparticles that can also lead to creation of magnetic moment. In fact, we have observed room-temperature ferromagnetism in CdS and CdSe nanoparticles. Similarly, room-temperature ferromagnetism and superconducting diamagnetism are observed in nanoparticles of δ-MoN as shown in Fig. 6. These results demonstrate that ferromagnetism may be a universal feature of nanoparticles of most compounds.

## 4. Conclusions

Room-temperature ferromagnetism is observed in nanoparticles of superconducting YBCO as well as metal nitrides. It requires further studies to explore the presence or absence of ferromagnetism in the superconducting state. The observation of room-temperature ferromagnetism in a variety of nanoparticles such as oxides, nitrides and chalcogenides suggests that ferromagnetism may be a universal feature of nanoparticles of most compounds.

## 5. Acknowledgement
We thank C. Madhu for his assistance in preparing some of the samples.6

**Figure Captions**

1. FESEM image of YBCO nanoparticles. Particle size ranges from 100 – 200 nm.

2. Volume susceptibility as a function of temperature showing the superconducting transition around 91 K. Note the absence of irreversibility between ZFC and FC curves down to 40 K. Inset shows the M(H) curve at 5 K.

3. M(H) data of YBCO nanoparticles at 300 K and 91 K showing the ferromagnetic behavior. M(H) data of bulk YBCO is also shown for comparison. Inset shows the hysteresis at 90 K which is typical of a superconductor.

4. M(H) data at 300 K for NbN nanoparticles (30 nm) showing ferromagnetic behavior. Inset at the bottom shows the superconducting diamagnetic onset around 6 K and the top inset shows the hysteresis in the superconducting state.

5. M(H) data at 300 K for NbN particles with an average size of ~ 200 nm showing a paramagnetic behavior. Hysteresis at 3.3 K shown in the top inset is typical of a bulk superconductor. The diamagnetic transition is shown in the bottom inset.

6. M(H) data at 300 K for δ-MoN. Inset shows the diamagnetic response under the applied field of 5 Oe.

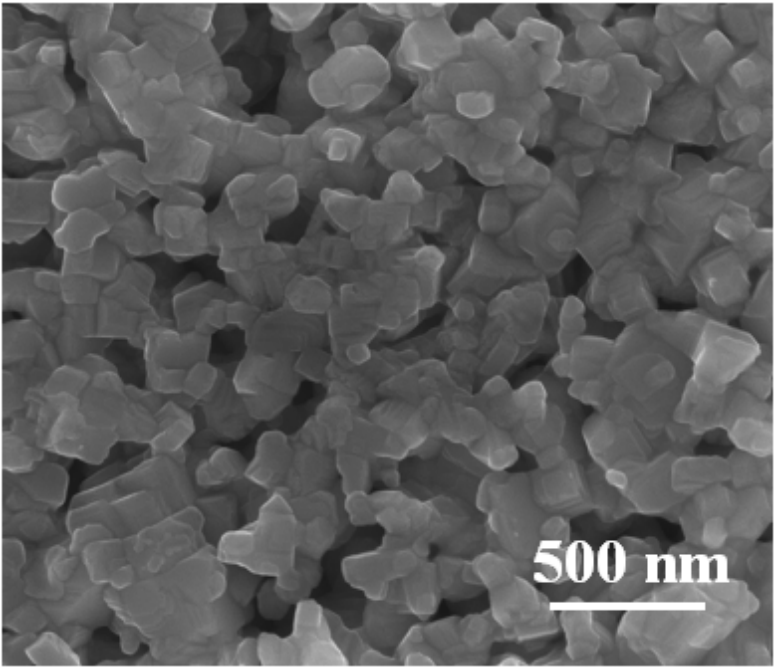

Fig1

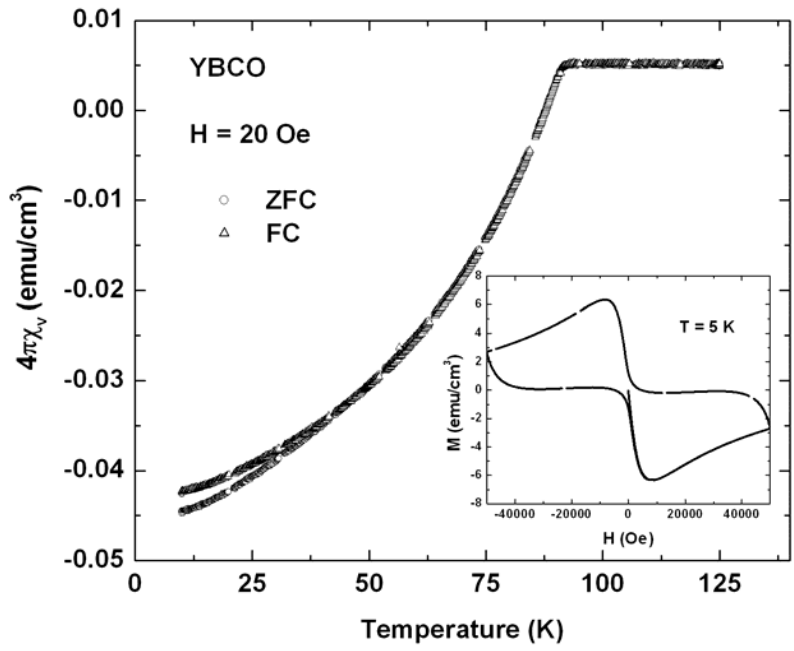

Fig2

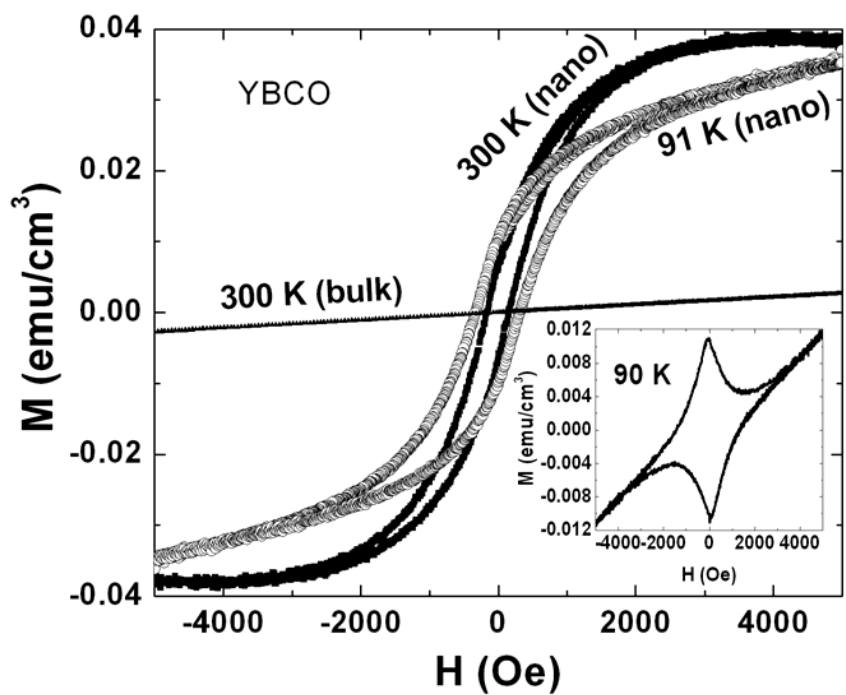

Fig3

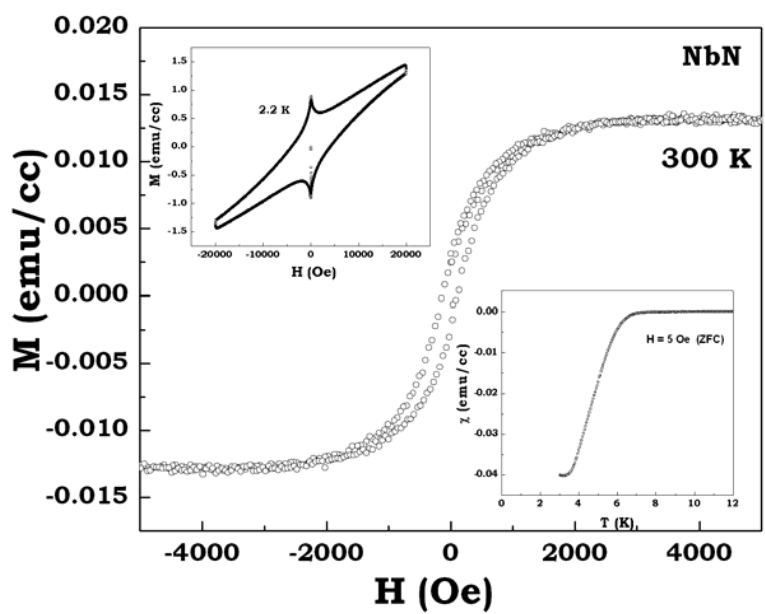

Fig4

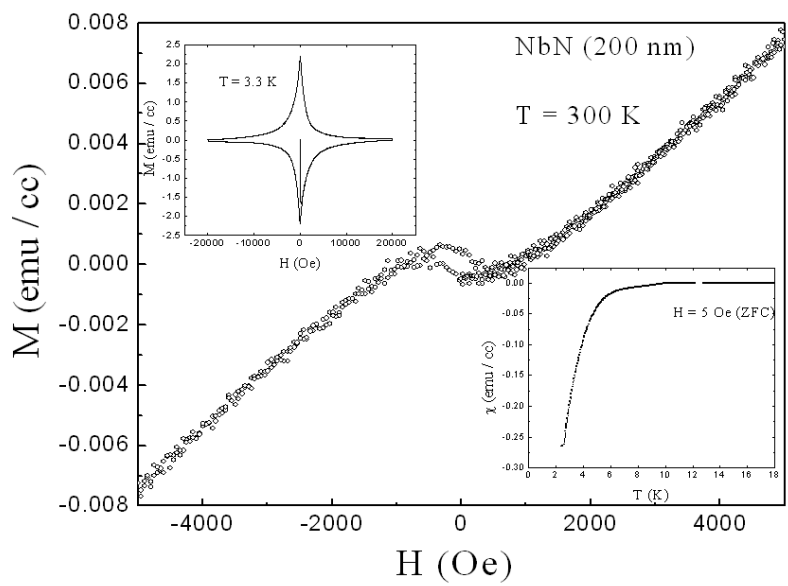

Fig5

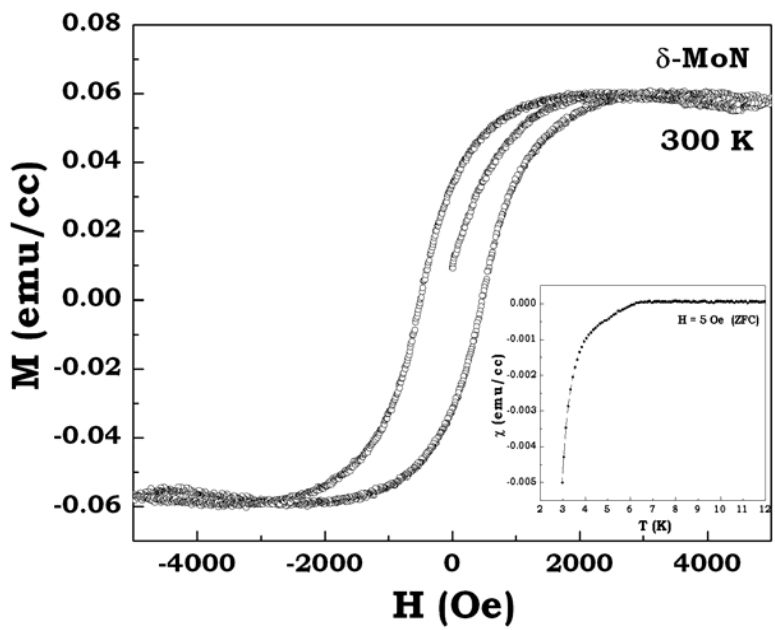

Fig6